\documentclass[a4paper,aps,prd,10pt,preprintnumbers,twocolumn,superscriptaddress,nofootinbib,amsmath,amssymb]{revtex4-1}
\usepackage{graphicx}
\usepackage[utf8]{inputenc}
\usepackage[T1]{fontenc}
\usepackage{cmap}
\usepackage{hyperref}
\usepackage{subfigure}
\usepackage{multirow}
\usepackage[toc,page]{appendix}
\DeclareMathAlphabet{\pazocal}{OMS}{zplm}{m}{n}

\def\imo{i}

\begin{document}
\title{Quasinormal modes, scattering and Hawking radiation in the vicinity of Einstein-dilaton-Gauss-Bonnet black hole}
\author{R. A. Konoplya}
\email{roman.konoplya@gmail.com}
\affiliation{Institute of Physics and Research Centre of Theoretical Physics and Astrophysics, Faculty of Philosophy and Science, Silesian University in Opava, CZ-746 01 Opava, Czech Republic}
\affiliation{Peoples Friendship University of Russia (RUDN University), 6 Miklukho-Maklaya Street, Moscow 117198, Russian Federation}
\author{A. F. Zinhailo}\email{F170631@fpf.slu.cz}
\affiliation{Institute of Physics and Research Centre of Theoretical Physics and Astrophysics, Faculty of Philosophy and Science, Silesian University in Opava, CZ-746 01 Opava, Czech Republic}
\author{Z. Stuchlík}\email{zdenek.stuchlik@fpf.slu.cz}
\affiliation{Institute of Physics and Research Centre of Theoretical Physics and Astrophysics, Faculty of Philosophy and Science, Silesian University in Opava, CZ-746 01 Opava, Czech Republic}
\begin{abstract}
Classical (quasinormal) and quantum (Hawking) radiations are investigated for test fields in the background of a four dimensional, spherically symmetric and asymptotically flat black hole in the Einstein-dilaton-Gauss-Bonnet (EdGB) theory. The geometry of the EdGB black hole deviates from the Schwarzschild geometry only slightly.  Therefore, here we observe that the quasinormal spectrum also deviates from its Schwarzschild limit at most moderately, allowing for a $9\%$  decrease in the damping rate and up to a $6\%$ decrease in the real oscillation frequency. However, the intensity of Hawking radiation of an electromagnetic field turned out to be much more sensitive characteristic than its quasinormal spectrum, allowing for a $54\%$  increase of the energy emission rate. The analytical formula for the eikonal regime of quasinormal modes is derived for test fields and it is shown that  the correspondence between the eikonal quasinormal modes and null geodesics is indeed fulfilled for test fields, but is not expected for the gravitational one.
\end{abstract}
\maketitle

\section{Introduction}

The Einstein theory of gravity, being consistent with recent observations of gravitational waves \cite{TheLIGOScientific:2016src}, nevertheless, leaves a number of fundamental questions open. These are construction of a non-contradicting quantum gravity, the nature of singularities, dark energy/dark matter problems and others. In addition, even the observations of gravitational waves leaves the window for alternative theories of gravity open \cite{Konoplya:2016pmh,Wei:2018aft,Berti:2018vdi}. Therefore, recently there has been a revival of interest to alternative theories of gravity, which have the same post-Newtonian behavior as the Einsteinian one, but different features in the strong field regime. Many of these theories include corrections in the form of higher orders in curvature and/or various scalar fields. Apparently, one of the most motivated of such approaches is the Einstein-dilaton-Gauss-Bonnet theory, which can be considered as an effective theory inspired by the low-energy limit of string theory \cite{string}, allowing one to test quantum corrections to General Relativity \cite{low-energy}. The theory consists of the Einstein action and the Gauss-Bonnet term, which is quadratic in curvature and coupled to the scalar (dilaton) field. In four-dimensional spacetimes, the Gauss-Bonnet term alone represents the full divergence and, therefore, does not contribute to the resulting equations of motion. This is not so, when the Gauss-Bonnet term is coupled to the dilaton.

The exact solution, describing a static, spherically symmetric and asymptotically flat black hole in the EdGB theory was found numerically in \cite{Kanti-metric}. Analytical approximation for this numerical solution has been obtained in \cite{Kokkotas:2017ymc}. The numerical spherical solution \cite{Kanti-metric} was further extended, also numerically, to the case of rotating black hole \cite{6}. There is also a perturbative solution for the EdGB black hole in terms of the rotation and coupling constant parameters \cite{7,8}. Various properties and potentially observable physical quantities have been recently considered for these black hole solutions. The reflection spectrum of accreting black holes has been studied in \cite{13,Nampalliwar:2018iru}, while its quasi-period oscillations were studied in \cite{14}. The shadows cast by the black hole were considered first in \cite{15}, and then in \cite{16}. The gravitational quasinormal modes, though  only partially (for some type of perturbations), were calculated in \cite{17,18,19}.

At the same time, no analysis of behavior of test fields in the vicinity of EdGB black holes were fulfilled, except for the analysis of separation of variables for the Klein-Gordon equation in the rotating black hole spacetime \cite{Konoplya:2018arm}. Calculations of energy emission rates of Hawking radiation in the vicinity of the Einstein-dilaton-Gauss-Bonnet black hole could possibly shed light on the influence of  quantum corrections upon black-hole evaporation, at least at the stage when the the mass of the black hole is several orders larger than the Planck mass and the full quantum gravity description can still be avoided. To the best of our knowledge, no such estimations for the evaporation of EdGB black hole have been done so far.

Usually, calculation of quasinormal modes of test fields is less interesting and motivated problem than that for the gravitational perturbations, because the latter one represents the fingerprints of gravitational waves. However, in our opinion the case of the Gauss-Bonnet term is interesting, because of the two reasons. The first one is related to the observation that the Gauss-Bonnet term alone in the higher dimensional theories makes the intensity of Hawking radiation much weaker even when the Gauss-Bonnet coupling constant is relatively small \cite{Konoplya:2010vz,Rizzo:2006uz}, that is, when the deviation from the D-dimensional Schwarzschild (Tangherlini) geometry is relatively small. Then, it would be natural to learn whether the same effect of strong suppression of Hawking radiation takes place in the four-dimensional case. The other reason for our study is related to the so called correspondence between high frequency (eikonal) quasinormal modes of a static, spherically symmetric black hole and the null geodesics in its background \cite{Cardoso:2008bp}.  Recently, it has been shown that the above correspondence is indeed guaranteed, but only for test fields, and not necessarily for the gravitational perturbations of the black-hole spacetime \cite{Konoplya:2017wot}, and a counter-example was given in the form of $D>4$-dimensional asymptotically flat Einstein-Gauss-Bonnet black hole. Therefore, it would be reasonable to compare the quasinormal modes of test fields with those for gravitational perturbations of the Einstein-dilaton-Gauss-Bonnet black hole and see whether the claimed correspondence is violated in a four-dimensional spacetime as well. The other issue which we would like to understand here is whether there are infinitely long lived modes of a massive test field, called quasi-resonances \cite{Ohashi:2004wr,Konoplya:2004wg,Konoplya:2005hr,Konoplya:2006br}, when the Gauss-Bonnet coupling is turned on.

Having all of the above motivations in mind we will compute here quasinormal modes of a test scalar and Maxwell fields and estimate the intensity of Hawking radiation for the Einstein-dilaton-Gauss-Bonnet black hole. We shall show the quantum (Hawking) radiation is much more sensitive to the relatively small dilaton-Gauss-Bonnet corrections of the Schwarzschild geometry than the classical radiation dominated by the quasinormal modes.

Our paper is organized as follows. Sec. II outlines the main features of the theory and parameterized black-hole metric under consideration. Sec. III is devoted to calculations of quasinormal modes with the help of the 6th order WKB method and time-domain integration. Sec. IV deduces the analytical formula for quasinormal modes in the eikonal regime and discusses the correspondence between the modes and null geodesics. In Sec. V we analyze the scattering properties and Hawking radiation of the electromagnetic field. Finally, we summarize the obtained results and mention open questions.

\section{The parameterized Einstein-dilaton-Gauss-Bonnet black hole metric}

The Lagrangian of the dilaton gravity with a Gauss Bonnet term is
\begin{eqnarray}
{\cal L}&=&\frac{1}{2}R - \frac{1}{4} \partial_\mu \phi \partial^\mu \phi \\\nonumber&&+ \frac{\alpha '}{8g^2} e^{\phi }\left(R_{\mu\nu\rho\sigma}R^{\mu\nu\rho\sigma} - 4 R_{\mu\nu}R^{\mu\nu} + R^2\right),
\end{eqnarray}
where $\alpha '$ is the Regge slope and $g$ is the gauge coupling constant.
In the general case the metric for a spherically symmetric black hole can be written in the form:
%
\begin{eqnarray}\label{metric}
ds^2 &=& -e^{\mu}dt^2+e^{\nu}{dr^2}+r^2 (\sin^2 \theta d\phi^2+d\theta^2),
\end{eqnarray}
where the analytical approximations of the numerical solution \cite{Kanti-metric} for functions $e^{\mu (r)}$ and $e^{\nu (r)}$ were found in \cite{Kokkotas:2017ymc} and are written down here up to the second order in Appendix I.

For convenience we fix $r_0=1$ and measure the radial coordinate in the units of the event horizon radius, so that the family of the EdGB black hole solutions can be parameterized via the following dimensionless parameter
\begin{equation}\label{fampar}
p\equiv6e^{2\phi_0}=\frac{6\alpha'^2}{g^4r_0^4}e^{2(\phi_0-\phi_{\infty})}\, , \qquad 0\leq p<1 \,.
\end{equation}
The limit $p=0$ corresponds to the Schwarzschild black hole; $\phi_0$ is the value of dilaton at the event horizon $r_0$. Owing to the above dilaton shifting, the shifted dilaton function goes to minus infinity  when $p \rightarrow 0$. This choice does not cause any problems for our purposes, because $e^{\phi(r)}$ remains always finite. It has been recently shown in \cite{Blazquez-Salcedo:2017txk} that black hole with $p\gtrsim0.97$ are linearly unstable against gravitational perturbations at lower multipole numbers $\ell$. Therefore, here we shall consider $p=0.97$ as the limiting case. Let us notice that the Gauss-Bonnet term at some critical value of $p$ might also produce  the so-called eikonal instability (see, for example,  \cite{Gleiser:2005ra,Takahashi:2012np,Takahashi:2012np,Konoplya:2017lhs,Konoplya:2017zwo} and references therein), which occurs at high multipole numbers $\ell$ and which was not analyzed for the case of the EdGB black hole in the literature so far. Recent studies of instabilities of wormhole solutions in the EdGB theory show agreement between the linear \cite{Cuyubamba:2018jdl} and nonlinear \cite{Shinkai:2017xkx} perturbations and come to the same conclusion on the instability of wormholes at whatever small value of the coupling constant.
\vspace{2mm}

\begin{figure}[ht]
\vspace{-4ex} \centering \subfigure[]{\includegraphics[width=0.8\linewidth]{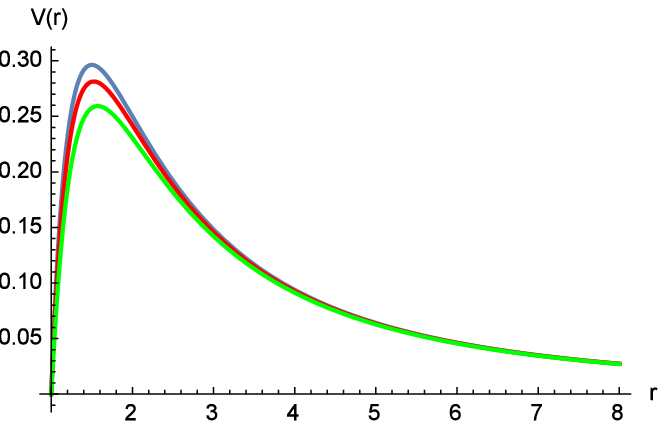} \label{fig:Velm_a} }
\hspace{4ex}
\subfigure[]{
\includegraphics[width=0.8\linewidth]{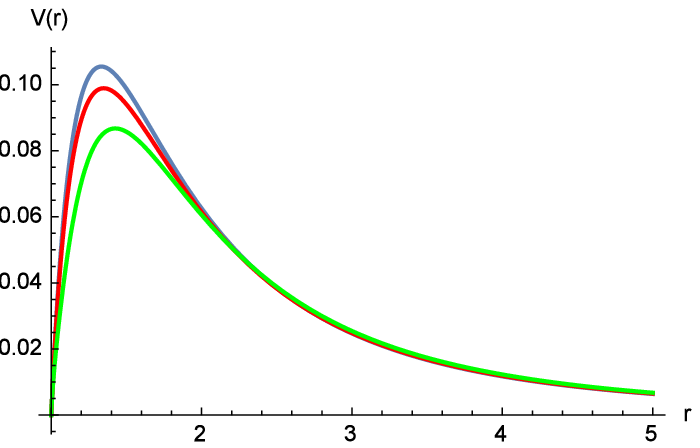} \label{fig:Vsc_b} }
\caption{ Dependence of the effective potentials $V(r)$ with $r_0=1$; blue line is $p=0$, red line is $p=0.4$ and green line is $p=0.97$:
 \subref{fig:Velm_a} $s = 1$, $\ell=1$;
 \subref{fig:Vsc_b} $s = 0$, $\ell=0$.}
 \label{ris:potentials}
\end{figure}

\section{Quasi-normal modes of test fields for the EdGB black hole}

 Massive scalar field is described by the general relativistic Klein-Gordon equation:
\begin{equation}\label{gensc}
\dfrac{1}{\sqrt{-g}} \partial_\mu(\sqrt{-g} g^{\mu\nu} \partial_\nu\Phi)-m^2 \Phi = 0,
\end{equation}
while an electromagnetic field obeys the general covariant Maxwell equations:
\begin{equation}\label{genelm}
\dfrac{1}{\sqrt{-g}} \partial_\mu(F_{\rho\sigma}g^{\rho\nu}g^{\sigma\mu}\sqrt{-g})=0.
\end{equation}
Here $A_{\mu}$ is a vector potential and $F_{\rho\sigma}=\partial_\rho A^\sigma - \partial_\sigma A^\rho$. In order to provide the separation of variables, the function $\Phi$ for the scalar field and a gauge invariant combination for the electromagnetic one can be expressed in terms of the spherical harmonics:
\begin{equation} \label{psi}
\Phi(t,r,y,\phi)~or~gauge~inv.~comb. \sim e^{-\imo\omega t} Y_l(\theta,\phi).
\end{equation}

After separation of angular variables, the wave equation can be represented in the following general form (see, for instance \cite{Konoplya:2006rv,Zinhailo:2018ska} and references therein):
\begin{equation}  \label{klein-Gordon}
\dfrac{d^2 \Psi}{dr_*^2}+(\omega^2-V_{i}(r))\Psi=0,
\end{equation}
where the relation
$$dr_*=\sqrt{e^{\nu-\mu}}dr$$
defines the ``tortoise coordinate'' $r_*$.

The effective potentials of test scalar ($i=s$) and electromagnetic ($i=e$) fields in the general background (\ref{metric}) can be written in the following forms:
%
\begin{subequations}\label{potentials}
\begin{eqnarray}\label{empotential}
V_{s} = \frac{e^{\nu} (e^{\mu})'- e^{\mu}(e^{\nu})'}{2 r e^{2 \nu}} +e^{\mu}\left(\frac{\ell(\ell+1)}{r^2} + m^2\right),
\end{eqnarray}
\begin{eqnarray}\label{scalarpotential}
V_{e}=\dfrac{\ell(\ell+1)e^{\mu}}{r^2}.
\end{eqnarray}
\end{subequations}
%
The effective potentials have the form of positive definite potential barriers decaying on the event horizon and at infinity (see figs. (\ref{fig:Velm_a},\ref{fig:Vsc_b})). It can be seen that the potential barrier becomes lower when the dilaton-GB term is on.


Quasinormal modes are eigenvalues of the above wave-like equation (\ref{klein-Gordon}), which can be written in the form
$$\omega = Re(\omega) + i Im(\omega),$$
and satisfy the following boundary conditions:
\begin{equation}
\Psi \sim \pm e^{\pm i \omega r^{*}}, \quad r^{*} \pm \infty.
\end{equation}
Here we imply that the negative $Im(\omega)$ corresponds to damping. The above boundary conditions mean that the waves are purely outgoing at infinity and purely incoming on the event horizon, i.e. no waves, coming from either the horizon or infinity, are allowed.
These boundary conditions represent the response of a black hole to a momentary perturbation, when the source of perturbation stopped acting \cite{reviews}.

For computation of quasinormal modes we will use the 6th order WKB formula \cite{WKBorder}. \textbf{It} is based on the WKB expansion of the wave function at both infinities (the event horizon and spacial infinity) which are matched with the Taylor expansion near the peak of the effective potential. The WKB approach in this form implies existence of two turning points and monotonic decay of the effective potentials along both infinities
\begin{equation}\label{WKB}
	\frac{i Q_{0}}{\sqrt{2 Q_{0}''}} - \sum_{i=2}^{i=p}
		\Lambda_{i} = n+\frac{1}{2},\qquad n=0,1,2\ldots,
\end{equation}
where the correction terms $\Lambda_{i}$ were obtained in \cite{WKBorder} for various orders up to the sixth. For \textbf{a} massive field it is sufficiently accurate only when the mass term $m^2$ is not large \cite{Konoplya:2017tvu}, and the higher $\ell$, the larger $m^2$ can be calculated relatively accurately. Although the WKB series converges only asymptotically, in the majority of cases the difference between the results obtained at higher and lower WKB orders gives an idea of how large is the expected error of the WKB approximation. Testing of the tenth order WKB formula \cite{Matyjasek:2017psv} shows that  for lower multipoles, in cases the 6th order is not optimal, the tenth order formula gives only a very small correction to the 6th order results \cite{Zhidenko-private}. Therefore, we think that the usage of the 6th order WKB formula is the most efficient for our purposes.

\vspace{2mm}
\begin{figure}[ht]
\vspace{-4ex} \centering \subfigure[]{
\includegraphics[width=0.7\linewidth]{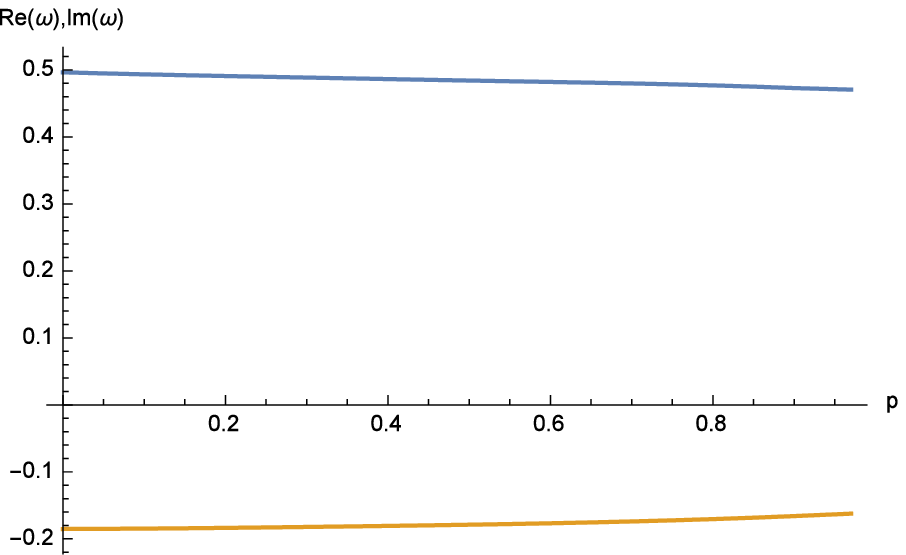} \label{fig:electromagnetic_a} }
\hspace{4ex}
\subfigure[]{
\includegraphics[width=0.7\linewidth]{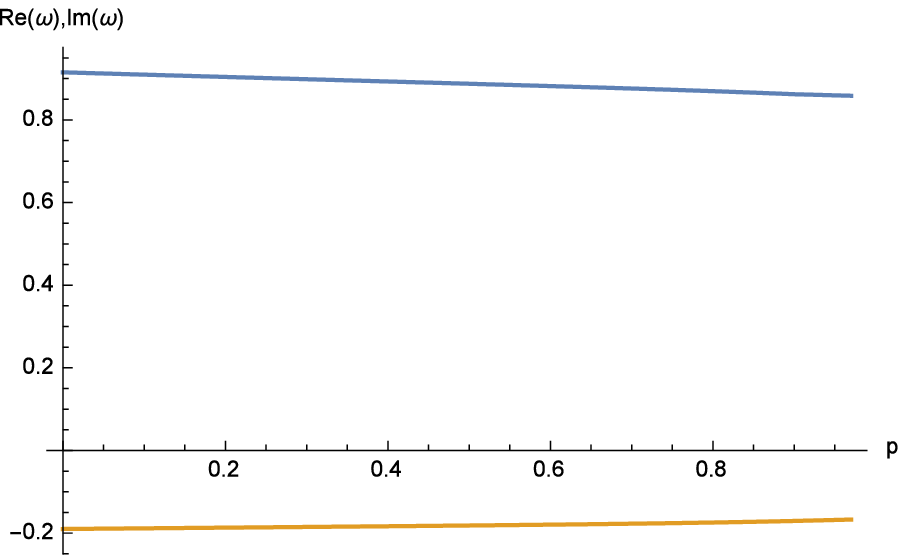} \label{fig:electromagnetic_b} }
\hspace{4ex}
\subfigure[]{ \includegraphics[width=0.7\linewidth]{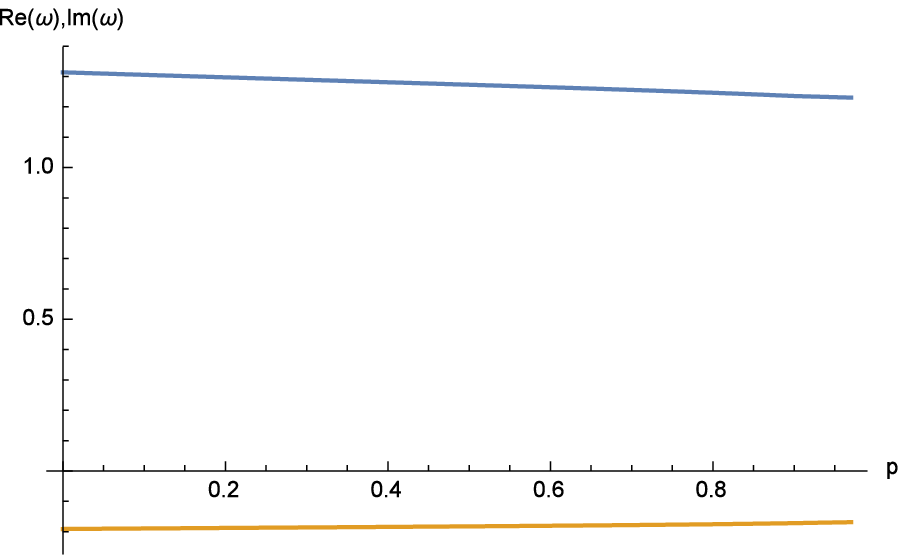} \label{fig:electromagnetic_c} }
\caption{The fundamental quasinormal mode ($n=0$) for the electromagnetic field ($s=1$), blue line is real part of frequencies, red line is imaginary part:
 \subref{fig:electromagnetic_a} $\ell = 1$;
 \subref{fig:electromagnetic_b} $\ell = 2$;
 \subref{fig:electromagnetic_c} $\ell = 3$.}
 \label{ris:one}
\end{figure}
\begin{figure}[ht]
\vspace{-4ex} \centering \subfigure[]{
\includegraphics[width=0.7\linewidth]{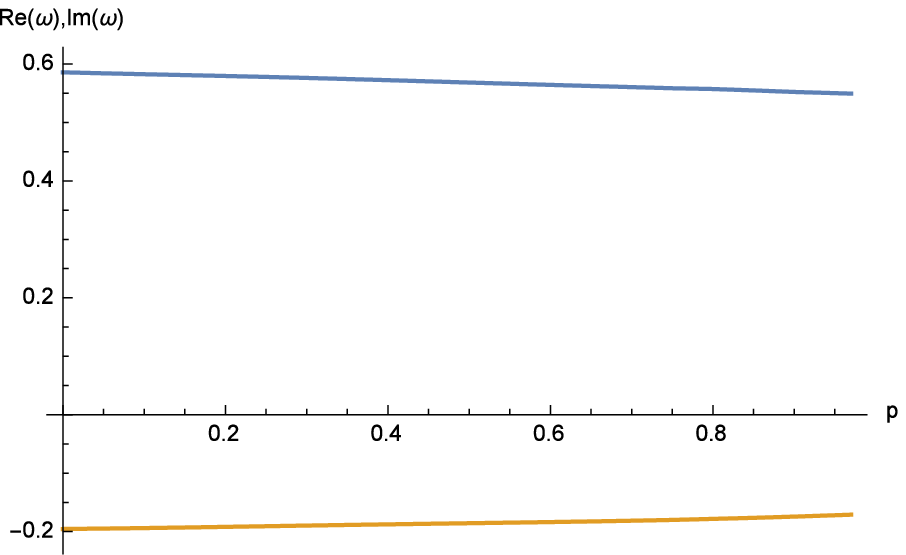} \label{fig:scalar_b} }
\hspace{4ex}
\subfigure[]{ \includegraphics[width=0.7\linewidth]{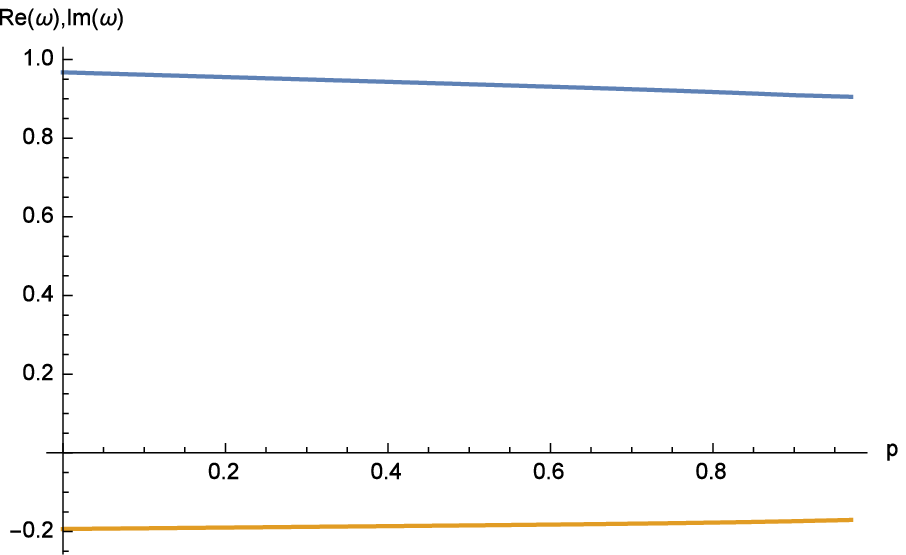} \label{fig:scalar_c} }
\caption{The fundamental quasinormal mode ($n=0$) for the scalar field ($s=0$), blue line is real part of frequencies, red line is imaginary part:
 \subref{fig:scalar_b} $\ell = 1$;
 \subref{fig:scalar_c} $\ell = 2$.}
 \label{ris:two}
\end{figure}

The quasi-normal modes for  $\ell = 1, 2$ for the scalar field ($s = 0$) and $\ell = 1, 2, 3$ for the electromagnetic field ($s = 1$) are presented on figs. (\ref{ris:one}, \ref{ris:two}). Variations of the imaginary and real parts of $\omega$ as functions of $p$  for various $\ell$ can be very well approximated by linear laws for the electromagnetic
\begin{subequations} \label{elmline}
\begin{eqnarray}
Re(\omega_{s=1,\ell=1})\approx0.496 + 0.024 p,
\nonumber\\
Im(\omega_{s=1,\ell=1})\approx-0.188 + 0.020 p;\\
Re(\omega_{s=1,\ell=2})\approx0.915 - 0.057 p,
\nonumber\\
Im(\omega_{s=1,\ell=2})\approx-0.190 + 0.001 p;\\
Re(\omega_{s=1,\ell=3})\approx1.315 - 0.008 p,
\nonumber\\
Im(\omega_{s=1,\ell=3})\approx-0.192 + 0.020 p.
\end{eqnarray}
\end{subequations}
and  scalar
\begin{subequations} \label{elmline}
\begin{eqnarray}
Re(\omega_{s=0,\ell=1})\approx0.587 - 0.037 p,
\nonumber\\
Im(\omega_{s=0,\ell=1})\approx-0.196 + 0.022 p;\\
Re(\omega_{s=0,\ell=2})\approx0.968 - 0.063 p,
\nonumber\\
Im(\omega_{s=0,\ell=2})\approx-0.194 + 0.021 p.
\end{eqnarray}
\end{subequations}
fields. These fits were obtained with the \emph{FindFormula} built-in function of \emph{Mathematica} when $p$ changes from $0$ to $0.9$. The exception is the case $\ell=n=0$ of the scalar field, for which the WKB method gives a big error even in the Schwarzschild limit and, therefore, cannot be trusted, especially for large values of $p$. From the above plots and fits one can see that the quasinormal modes of EdGB black hole have smaller damping rates and oscillation frequencies than those of the Schwarzschild black hole. At the same time, we can see that the effect of the Gauss-Bonnet coupling on quasinormal modes is not large, being about $6\%$ for $Re (\omega)$ and about $9\%$ for the $Im (\omega)$. Here, we had in mind that the continued fraction converges slowly near the extremal value $p=1$, so that only the values of $p$ which are far enough from the extremal case can be trusted. Another kind of error comes from usage of the WKB method, once $\ell$ is small. Therefore, in order to claim the above mentioned effect on real and imaginary parts, we considered values of $p$ such that:
\begin{enumerate}
\item the relative error coming from the analytical approximation of the metric is still well within of few tenth of a percent, that is the error is much smaller than the effect
\item the multipole number $\ell$ is high enough to neglect the error of the WKB formula.
\end{enumerate}
Let us notice that the above effect of up to $9\%$ for the damping rate of quasinormal mode is much larger than the other studied effects in the electromagnetic spectrum. Thus, the shadows of the EdGB black hole was shown to be changed by around one percent \cite{15,16} and a similar small change is observed in the iron line \cite{Nampalliwar:2018iru} in the context of X-ray reflection spectroscopy.

\begin{table}
\begin{tabular}{|c|c|}
  \hline
    \hline
    $p$ & $\omega$ \\
      \hline
        \hline
    $0$ & $0.221 - 0.210 i$ \\
      \hline
    $0.1$ & $0.2207 - 0.208 i$ \\
      \hline
    $0.5$ & $0.219 - 0.200 i$ \\
      \hline
    $0.9$ & $0.216 - 0.186 i$ \\
      \hline
    $0.97$ & $0.215 - 0.182 i$ \\
  \hline
    \hline
\end{tabular}
\caption{The $\ell=n=0$ mode of a scalar field perturbation obtained by the time-domain integration and extracted by the Prony method.}
\label{TableL0}
\end{table}

\begin{figure}
\resizebox{0.85\linewidth}{!}{\includegraphics*{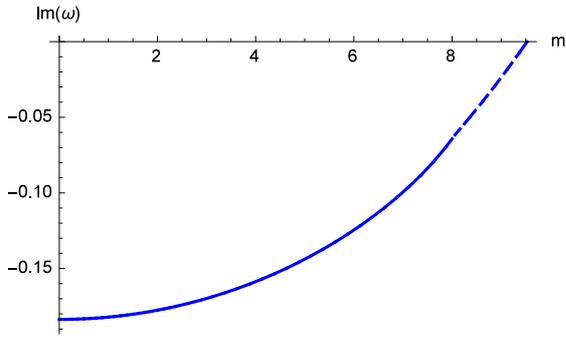}}
\caption{Imaginary part of the quasinormal modes for a massive scalar field $\ell =15$, $p=0.5$.}\label{L15Im}
\end{figure}

\begin{figure}
\resizebox{0.8\linewidth}{!}{\includegraphics*{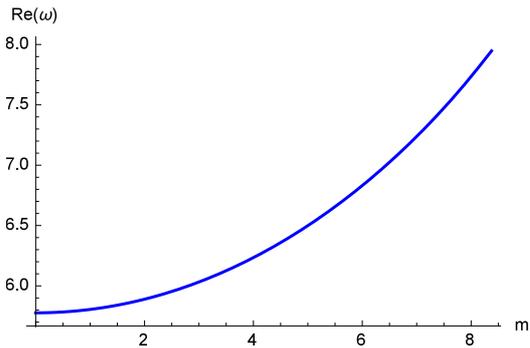}}
\caption{Real part of the quasinormal modes for a massive scalar field $\ell =15$, $p=0.5$.}\label{L15Re}
\end{figure}

It is known that massive test fields allow for infinitely long lived modes, called quasi-resonances, at some threshold  values of mass $m$ for the Schwarzschild and Kerr black holes \cite{Ohashi:2004wr,Konoplya:2004wg,Konoplya:2005hr,Konoplya:2006br}, as well as for some black hole solutions in alternative theories of gravity \cite{Zinhailo:2018ska}. From  figs. (\ref{L15Im}, \ref{L15Re}) for quasinormal modes of a massive scalar field at some fixed high $\ell$ we can see that there is a clear indication that quasi-resonances exist for the EdGB black hole as well. Even though the WKB formula \cite{WKBorder} cannot be used for evaluation of very long lived modes \cite{Konoplya:2017tvu}, the WKB method is quite accurate in the regime of high $\ell$ and the extrapolation to higher values of mass $m$ gives an evidence that quasi-resonances should exist for the EdGB black hole as well.

\begin{figure}
\resizebox{0.9\linewidth}{!}{\includegraphics*{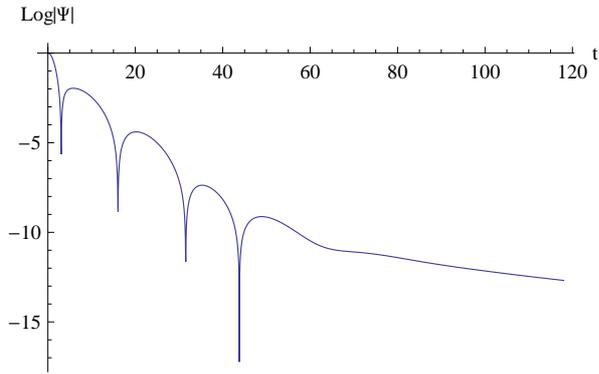}}
\caption{Time-domain profile for the scalar field's perturbation: $\ell=0$, $p=0.97$. }\label{TimeD}
\end{figure}

For the $\ell=n=0$ mode the error of the WKB method is usually larger than the effect, so that we have to use the time-domain integration of the perturbation equations instead. We will integrate the wavelike equation rewritten in terms of the light-cone variables $u = t - r_*$ and $v = t + r_*$. The appropriate discretization scheme was suggested in \cite{Gundlach:1993tp}:
\begin{eqnarray}\label{eq:scheme}
\Psi(N) &=& \Psi(W) + \Psi(E) -
\Psi(S)  \\
& &- \Delta^2\frac{V(W)\Psi(W) + V(E)\Psi(E)}{8} +
\mathcal{O}(\Delta^4)\ ,
\nonumber
\end{eqnarray}
where we used the following definitions for the points: $N = (u + \Delta, v + \Delta)$, $W = (u + \Delta, v)$, $E = (u, v + \Delta)$ and $S = (u,v)$. The initial data are specified on the null surfaces $u = u_{0}$ and $v = v_{0}$ From fig. (\ref{TimeD}) it can be seen that the $\ell=0$ mode is characterized by a relatively short period of quasinormal ringing. Therefore, it is difficult to extract values of frequencies with good accuracy, so that in the table \ref{TableL0} we write down only the three digits after point for $\omega$.

\section{Analytical formula for QNMs of test fields in the eikonal regime}

In the regime of high multipole numbers $\ell$ (eikonal) the behavior of test fields of different spin obey the same law in the dominant order. Here we will consider an electromagnetic field, governed by the effective potential (\ref{empotential}). In the eikonal regime it is sufficient to use the first order WKB formula:
\begin{equation}\label{wkb}
\omega=\sqrt{{V_0}-i \left(n+\frac{1}{2}\right) \sqrt{-2 {V_0''}}}.
\end{equation}
Here, $ n=0,1,2,..$ is a overtone number, ${V_0}$ is a value of $V(r)$ in $r_{max}$ and $V_0''$ is a second derivative in $r_{max}$. Making series expansion in terms of $1/\ell$ we find that the maximum of the effective potential occurs at larger distance from the black hole:
\begin{equation}\label{rmax}
r_{max}= \frac{3 r_0}{2} + 0.055 r_{0} p +  \mathcal{O}(p^2).
\end{equation}

Substituting the above value of $r_{max}$ (\ref{rmax}) into (\ref{wkb}) and expanding in terms of $1/\ell$, we find
\begin{equation} \label{w}
\omega=\frac{(1+2 \ell)(1 -0.065 p) - i (2 n+1)(1-0.094 p)}{3 \sqrt{3}r_0}
\end{equation}
In the limit $p \rightarrow 0$ this formula goes over into the well-known eikonal formula for the Schwarzschild black hole.

Following Cardoso et. al. \cite{Cardoso:2008bp}, one can see that the principal Lyapunov exponent for null geodesics around a static,  spherically symmetric metric (\ref{metric}) is
\begin{equation}\label{GenLyap}
\lambda = \frac{1}{\sqrt{2}}\sqrt{-\frac{r_c^2}{(e^{\mu})_{c}}\left(\frac{d^2}{dr_*^2}\frac{e^{\mu}}{r^2}\right)_{r=r_c}}.
\end{equation}
The coordinate angular velocity for the null geodesics is
\begin{equation}\label{angularvel}
\Omega_c = \frac{(e^{\frac{\mu}{2}})_{c}}{r_c},
\end{equation}
where $r_{c}$ is the radius of the circular null geodesics,  satisfying the equation
\begin{equation}\label{circulareq}
 2 (e^{\mu})_c=r_c (e^{\mu})'_{c}.
\end{equation}
Comparing the eikonal formula (\ref{w}) with the Lyapunov exponent and angular velocity, one can easily see that the correspondence is fulfilled here for the test fields. However, quasinormal modes of gravitational perturbations of the EdGB black hole should not obey the above correspondence, in a similar fashion with higher dimensional Einstein-Gauss-Bonnet solution \cite{Konoplya:2017wot}. Even though the accurate analysis of gravitational perturbations  is beyond the scope of our work, it can easily be noticed from \cite{Konoplya:2017wot} that the necessary condition for the validity of the null geodesics/QNMs correspondence is the requirement that the effective potential in the eikonal regime have a ``good''. WKB form:
\begin{equation}\label{goodWKB}
V = e^{\mu} \frac{\ell (\ell +1)}{r^{2}} + \mathcal{O}(1/\ell).
\end{equation}
This is not so for gravitational perturbations of the higher dimensional Einstein-Gauss-Bonnet black hole, which is, thus, the counterexample to the correspondence in $D>4$-dimensional spacetimes. Analysis of gravitational perturbations of EdGB solution shows that the gravitational perturbations obey the following set of chained equations (see eqs. (16) in \cite{Blazquez-Salcedo:2017txk}):
$$
\frac{d \Psi_{(i)}}{dr} = U_{(i)} \Psi_{(i)} =0
$$
where, here $(i)$ stands for ``axial''. and ``polar''. types of perturbations and $\Psi$ is a two-dimensional vector for axial and four-dimensional vector for polar perturbations. The asymptotic behavior of these equations on both extremities, explicitly written in eqs. (17-30) of \cite{Blazquez-Salcedo:2017txk} implies that the necessary form of the effective potential (\ref{goodWKB}) is violated for the EdGB case.
Thus, the Einstein-dilaton-Gauss-Bonnet case is the four-dimensional counterexample of the correspondence, which, at the same time, confirms the claim  that the correspondence is always guaranteed for test fields \cite{Konoplya:2017wot}.

\section{Gray-body factors}

The intensity of Hawking radiation is always partially suppressed  by the effective potential surrounding the black holes, because the part of the total flow of particles emitted by the black hole is reflected back to the event horizon.
In order to estimate the number of particles reflected by the effective potential we need first to find the gray-body factors, that is, to solve the classical scattering problem.

We shall consider the wave equation (\ref{klein-Gordon}) with the boundary conditions allowing for incoming waves from infinity. Owing to the symmetry of the scattering properties this is identical to the scattering of a wave coming from the horizon. The scattering boundary conditions for (\ref{klein-Gordon}) have the following form
\begin{equation}\label{BC}
\begin{array}{ccll}
    \Psi &=& e^{-i\omega r_*} + R e^{i\omega r_*},& r_* \rightarrow +\infty, \\
    \Psi &=& T e^{-i\omega r_*},& r_* \rightarrow -\infty, \\
\end{array}%
\end{equation}
where $R$ and $T$ are the reflection and transmission coefficients.

\begin{figure}[ht]
\vspace{-4ex} \centering \subfigure[]{
\includegraphics[width=0.7\linewidth]{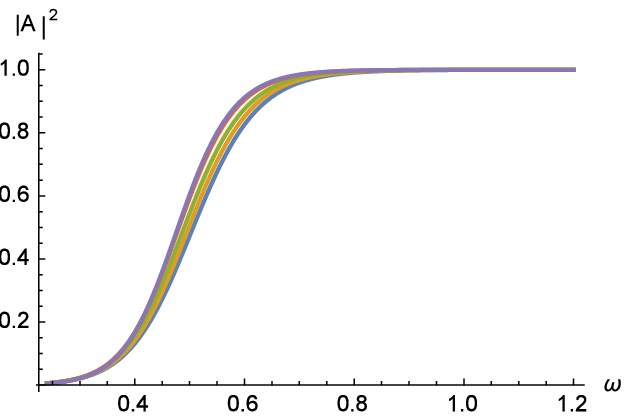} \label{fig:GBFactor_a} }
\hspace{4ex}
\subfigure[]{
\includegraphics[width=0.7\linewidth]{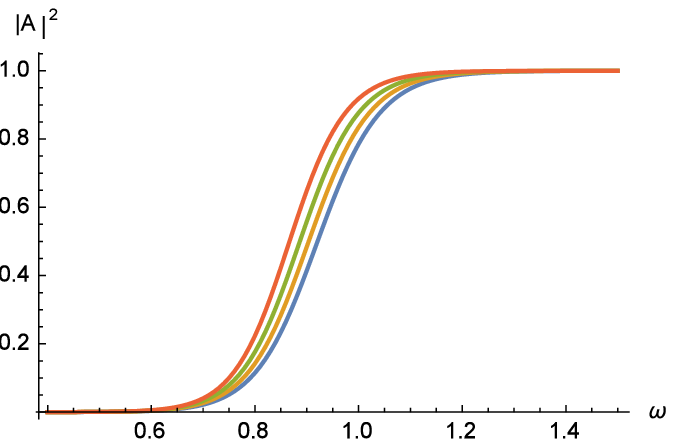} \label{fig:GBFactor_b} }
\hspace{4ex}
\subfigure[]{ \includegraphics[width=0.7\linewidth]{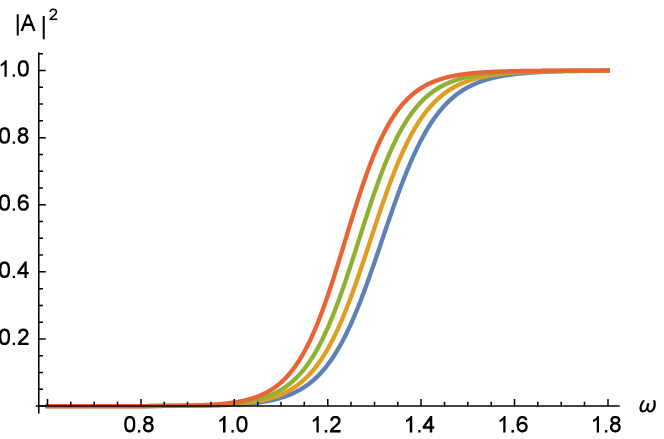} \label{fig:GBFactor_c} }
\caption{The gray-body factor dependence in $\omega$, for the electromagnetic field ($s=1$), blue line corresponds to $p=0$, it is Schwarzschild limit, orange line corresponds to $p=0.3$, green line corresponds to $p=0.6$, red line is a maximum value of $p=0.9$:
 \subref{fig:GBFactor_a} $l = 1$;
 \subref{fig:GBFactor_b} $l = 2$;
 \subref{fig:GBFactor_c} $l = 3$.}
 \label{ris:three}
 \end{figure}

\par
The effective potential has the form of the potential barrier which monotonically decreases at both infinities, so that the WKB approach \cite{WKBorder} can be applied for finding $R$ and $T$. Since $\omega^2$ is real, the first order WKB values for $R$ and $T$ will be real \cite{WKBorder} and
\begin{equation}\label{1}
\left|T\right|^2 + \left|R\right|^2 = 1.
\end{equation}
Once the reflection coefficient is calculated, we can find the transmission coefficient for each multipole number $\ell$
\begin{equation}
\left|{\pazocal
A}_{\ell}\right|^2=1-\left|R_{\ell}\right|^2=\left|T_{\ell}\right|^2.
\end{equation}
Various methods for computation of the transmission and reflection coefficients exist in the literature.
For quick and relatively accurate evaluation of the transmission and reflection coefficients we used the 6th order WKB formula \cite{WKBorder}.
The above formula does not work well when $\omega$ is very small, but fortunately, this corresponds to almost complete reflection of the waves and does not contribute noticeably into the total energy emission rate. In order to study contributions of particles at very small frequencies one can apply the first order WKB formula which will give more accurate results in this regime. This situation occurs because the WKB series converges only asymptotically and does not guarantee convergence in each WKB order. According to \cite{WKBorder} the reflection coefficient can be expressed as follows:
\begin{equation}\label{moderate-omega-wkb}
R = (1 + e^{- 2 i \pi K})^{-\frac{1}{2}},
\end{equation}
where $K$ can be determined from the following equation:
\begin{equation}
K - i \frac{(\omega^2 - V_{0})}{\sqrt{-2 V_{0}^{\prime \prime}}} - \sum_{i=2}^{i=6} \Lambda_{i}(K) =0.
\end{equation}
Here $V_0$ is the maximum of the effective potential, $V_{0}^{\prime \prime}$ is the second derivative of the
effective potential in its maximum with respect to the tortoise coordinate, and $\Lambda_i$  are higher order WKB corrections which depend on up to $2i$th order derivatives of the effective potential at its maximum \cite{WKBorder} and $K$.
On figs. (\ref{fig:GBFactor_a}, \ref{fig:GBFactor_b}, \ref{fig:GBFactor_c}) we show dependence of the gray-body factor on $\omega$ for different values of $l$ for an electromagnetic field. There one can see that the gray-body factors of the EdGB black hole are larger than that of the Schwarzschild limit. From the above figs. (\ref{fig:GBFactor_a},\ref{fig:GBFactor_b},\ref{fig:GBFactor_c}) one can see that for larger $p$, the transmission rate is higher. This agrees with the fact that the pick of the effective potential becomes lower at larger $p$, so that the tunneling through the lower potential is easier.

\begin{figure}
\resizebox{0.8\linewidth}{!}{\includegraphics*{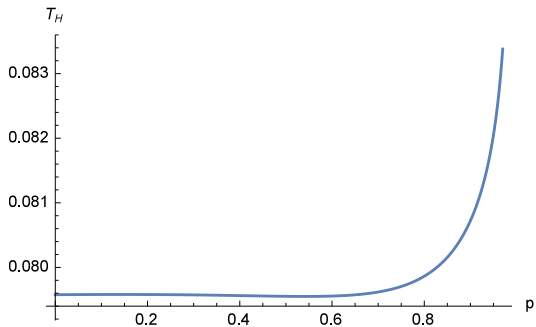}}
\caption{The Hawking temperature $T_H(p)$.}\label{TH}
\end{figure}

\vspace{3mm}

\section{Hawking radiation}

We will assume that the black hole is in the thermal equilibrium with its environment in the following sense: the temperature of the black hole does not change between emissions of two consequent particles. This implies that the system can be described by the canonical ensemble (see \cite{Kanti:2004nr} for a review). Therefore, the energy emission rate for Hawking radiation has the form \cite{Hawking:1974sw}:
\begin{align}\label{energy-emission-rate}
\frac{\text{d}E}{\text{d} t} = \sum_{l}^{} N_l \left| \pazocal{A}_l \right|^2 \frac{\omega}{\exp\left(\omega/T_\text{H}\right)-1} \frac{\text{d} \omega}{2 \pi},
\end{align}
were $T_H$ is the Hawking temperature, $A_l$ are the gray-body factors, and $N_l$ are the multiplicities, which only depend on the space-time dimension and $l$.
\begin{widetext}
\begin{table*}[ht]
\centering
\begin{tabular}{ | p{2cm} | p{2.5cm} | p{2.5cm} | p{2.5cm} | p{2.5cm} |}
\hline
\hline
$d E/d t$  & $p=0$ & $p=0.3$ & $p=0.9$ & $p=0.97$ \\
\hline
\hline
$\ell=1$    & $13.4034\cdot10^{-5}$ & $13.9278\cdot10^{-5}$  & $17.0736\cdot10^{-5}$ & $20.7412\cdot10^{-5}$ \\
$\ell=2$    & $2.6717\cdot10^{-6}$ & $3.0984\cdot10^{-6}$ & $4.8903\cdot10^{-6}$ & $6.9153\cdot10^{-6}$ \\
$\ell=3$    & $4.0262\cdot10^{-8}$ & $5.1084\cdot10^{-8}$ & $1.0359\cdot10^{-8}$ & $1.7117\cdot10^{-8}$ \\
\hline
$\sum\limits_{\ell=1}^3 (d E/d t)_{\ell} $  & $13.6746\cdot10^{-5}$ & $14.2428\cdot10^{-5}$ & $17.5730\cdot10^{-5}$ & $21.4499\cdot10^{-5}$ \\
\hline
\hline
\end{tabular}
\caption{\label{tab:table} The partial (for various $\ell$) and total energy emission rates of the electromagnetic field $d E/d t$ for different values  $p$.}
 \label{table1}
\end{table*}
\end{widetext}

\begin{figure}
\resizebox{0.8\linewidth}{!}{\includegraphics*{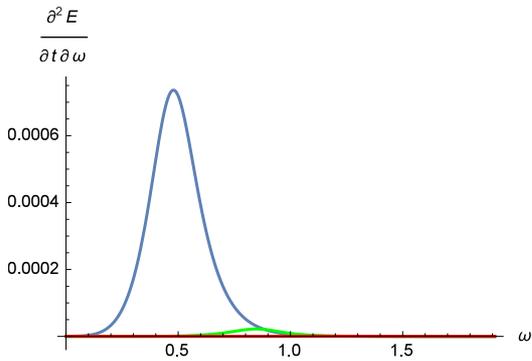}}
\caption{Energy emission rate of the electromagnetic field for different values $\ell$, $p=0.97$: blue line corresponds to $\ell=1$, green line corresponds to $\ell=2$, red line is $\ell=3$}\label{fig:qq}
\end{figure}

\begin{figure}
\resizebox{0.8\linewidth}{!}{\includegraphics*{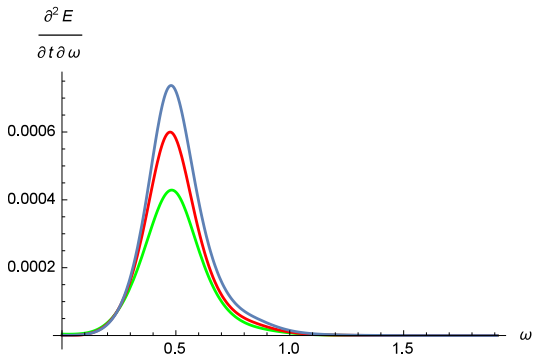}}
\caption{Total energy emission rate (after summation over all degrees of freedom and quantum numbers) of the electromagnetic field for $p =0$ (green), $0.9$ (red), and $0.97$ (blue)}\label{fig:pp}
\end{figure}

In the regime of small deviations from the Schwarzschild limit, $p \ll 1$, the Hawking temperature is
\begin{equation}
T_{H} = \frac{1}{4 \pi} + \frac{p}{7722 \pi} + \mathcal{O}(p^2).
\end{equation}
Dependence of the temperature on $p$ is shown on fig. (\ref{TH}) and the minimum ($T_{H} \approx  0.0796$) occurs at $p \approx  0.5411$.

From fig. (\ref{fig:qq}) one can see that the emission of electromagnetic radiation is highly dominated by the first multipole $\ell=1$, while contribution of the second multipole is small, while higher $\ell$ are almost negligible. Thus, the full picture of the electromagnetic radiation is very well represented by its $\ell=1$ and $\ell=2$ modes. From fig. (\ref{fig:pp})  it can be seen that the turning on of the dilaton-Gauss-Bonnet term enhances the intensity of Hawking radiation seemingly. At the same time, as the Hawking temperature increases by only about $4 \%$ from $p=0$ to $p=0.97$, the pick of the energy emission rate shifts to higher frequencies $\omega$ insignificantly.

The total energy emission rate of the electromagnetic field (see table \ref{table1}), i.e. after integrating over all the frequencies $\omega$ and multipole numbers $\ell$, is $13.7 \cdot 10^{-5}$ for $p=0$, $17.6 \cdot 10^{-5}$ for $p=0.9$ and $21.5 \cdot 10^{-5}$ for $p=0.97$. Thus, the maximal increase in the energy emission rate of Hawking radiation for an electromagnetic field reaches $54 \%$. This way, the Hawking radiation of an electromagnetic field is much more sensitive characteristic than classical radiation in the electromagnetic spectrum. In the Schwarzschild limit $p=0$ we reproduce the results of the well-known Page's work \cite{Page:1976df} after changing to the units $M=1/2$ ($r_{0}=1$): as the power scales as $\sim \hbar c^{6} G^{-2} M^{-2}$, one should divide  by $4$ the results obtained here in order to reproduce the Page's data.
\vspace{3mm}

\section{Conclusions}

In this work we filled the gap existing in investigation of radiation phenomena in the vicinity of the Einstein-dilaton-Gauss-Bonnet black hole.
Here we have shown that:
\begin{itemize}
\item As the geometry of the EdGB black hole is only slightly corrected by the Gauss-Bonnet-dilaton term, the quasinormal modes of test scalar and electromagnetic fields are only slightly  deviate from their Schwarzschild limits. Nevertheless, the damping rate of the quasinormal mode, allowing for $9$ percent deviation from the Schwarzschild limit, turns out to be more sensitive characteristic than some other observables in the electromagnetic spectrum, such as the size of the shadow or iron line  \cite{15,16,Nampalliwar:2018iru}. The latter usually give effect of not more than one percent.  The fundamental modes can be approximated very well by a linear fit in terms of the new parameter $p$ of the theory.
\item In the regime of high frequencies (eikonal) the analytical formula of quasinormal modes has been deduced. It generalizes that for the Schwarzschild black hole and confirm the QNM/null geodesics correspondence for test fields. However, it is argued that the correspondence should not work for the gravitational perturbations of EdGB black hole.
\item Massive fields in the vicinity of the EdGB black hole allow for infinitely long lived modes, called quasi-resonances.
\item The Hawking radiation is noticeably enhanced up to about $54\%$ when the dilaton-Gauss-Bonnet term is turned on.
\end{itemize}
\vspace{1mm}

One of the most appealing questions which was beyond our paper is the analysis of gravitational perturbations in order to see whether the eikonal instability occurs for that case as well. The appropriate Hawking radiation of gravitons could then be analyzed. However, in the region of black hole stability, the main features of the classical and quantum radiation considered here for test fields should also hold for the gravitational field, as it usually takes place for other stable black hole models.

\vspace{1mm}
\acknowledgments{
The authors acknowledge  the  support  of  the  grant  19-03950S of Czech Science Foundation ($GA\check{C}R$) and Alexander Zhidenko for useful discussions. R. K. would also like to thank the support of the COST Action CA16104 when visiting T\"ubingen University and the Mobility Grant of Silesian University in Opava at the initial stage of this work.
}

\vspace{1mm}

\begin{widetext}

\section{Appendix: The parameterized EdGB metric}\label{Appendix}

The  analytical approximations \cite{Kokkotas:2017ymc} for the metric functions $e^{\mu}$ and $e^{\nu}$ have the forms:
\begin{subequations}
\begin{eqnarray}
e^{\mu} &=&
[(r - r_{0}) (11528 (-338485+167871 p + 937132 p^2-1091895 p^3+325377 p^4) r^4+8 (263522875
\nonumber \\&&
+497564855 p-2160940683 p^2+1833700801 p^3-382791763 p^4-54635232 p^5+3579147 p^6) r^3 r_{0}
\nonumber \\&&
-124488 (-1+p)^2 p (-1310+1551 p-514 p^2+33 p^3) r^2 r_{0}^2+p (283646440-1112933120 p
\nonumber \\&&
+1868830098 p^2-1478746401 p^3+470844780 p^4-32741280 p^5) r r_{0}^3+1441 p (-234080+345600 p
\nonumber \\&&
-85004 p^2-36868 p^3+11115 p^4) r_{0}^4)]/[11528 (-1+p) (-5+3 p) r^4 ((-67697-74741 p
\nonumber \\&&
+108459 p^2) r+(36575+121424 p-124020 p^2) r_{0})],
\end{eqnarray}
\begin{eqnarray}
e^{\nu} &=&
[2882 (-1+ p) (-5+3 p) r^2 ((-67697-74741 p+108459 p^2) r+(36575+121424 p
\nonumber \\&&
-124020 p^2) r_{0}) (18 (-297882+533046 p-262075 p^2+24795 p^3) r^2+18 (223782-348366 p+110455 p^2
\nonumber \\&&
+16245 p^3) r r_{0}-95 p (-3640+8312 p-6075 p^2+1404 p^3) r_{0}^2)^2]/[81 (13-9 p)^2 (r-r_{0}) ((22914
\nonumber \\&&
-25140 p+2755 p^2) r+(-17214+14880 p+1805 p^2) r_{0})^2 (11528 (-338485+167871 p+937132 p^2
\nonumber \\&&
-1091895 p^3+325377 p^4) r^4+8 (263522875+497564855 p-2160940683 p^2+1833700801 p^3
\nonumber \\&&
-382791763 p^4-54635232 p^5+3579147 p^6) r^3 r_{0}-124488 (-1+p)^2 p (-1310+1551 p-514 p^2
\nonumber \\&&
+33 p^3) r^2 r_{0}^2+p (283646440-1112933120 p+1868830098 p^2-1478746401 p^3+470844780 p^4
\nonumber \\&&
-32741280 p^5) r r_{0}^3+1441 p (-234080+345600 p-85004 p^2-36868 p^3+11115 p^4) r_{0}^4)].
\end{eqnarray}
\end{subequations}

The above cumbersome expressions can be found in the form of Mathematica notebook in \text{https://arxiv.org/src/1705.09875v3/anc/Approximation.nb}.

\end{widetext}


\newpage
\begin{thebibliography}{80}
\bibitem{TheLIGOScientific:2016src}
  B.~P.~Abbott {\it et al.} [LIGO Scientific and Virgo Collaborations],
  Phys.\ Rev.\ Lett.\  {\bf 116}, no. 6, 061102 (2016)
  [arXiv:1602.03837 [gr-qc]];
  Phys.\ Rev.\ Lett.\  {\bf 116}, no. 22, 221101 (2016)
  [arXiv:1602.03841 [gr-qc]];
  Phys.\ Rev.\ Lett.\  {\bf 116}, no. 24, 241103 (2016)
  [arXiv:1606.04855 [gr-qc]].
\bibitem{Konoplya:2016pmh}
  R.~Konoplya and A.~Zhidenko,
  Phys.\ Lett.\ B {\bf 756}, 350 (2016)
  [arXiv:1602.04738 [gr-qc]].
\bibitem{Wei:2018aft}
  S.~W.~Wei and Y.~X.~Liu,
  Phys.\ Rev.\ D {\bf 98}, no. 2, 024042 (2018)
  [arXiv:1803.09530 [gr-qc]].
\bibitem{Berti:2018vdi}
  E.~Berti, K.~Yagi, H.~Yang and N.~Yunes,
  Gen.\ Rel.\ Grav.\  {\bf 50}, no. 5, 49 (2018)
  [arXiv:1801.03587 [gr-qc]].
\bibitem{string} J. Polchinski, String Theory,  (Cambridge University Press, Cambridge, 1998).
\bibitem{low-energy} C. G. . Callan, I. R. Klebanov and M. J. Perry, Nucl. Phys. B278 (1986) 78;D. J. Gross and E. Witten, Nucl. Phys. B 277 (1986) 1
\bibitem{Kanti-metric} P. Kanti, N. E. Mavromatos, J. Rizos, K. Tamvakis and E. Winstanley,  Phys. Rev. D54,  5049 (1996)
\bibitem{Kokkotas:2017ymc}
  K.~D.~Kokkotas, R.~A.~Konoplya and A.~Zhidenko,
  Phys.\ Rev.\ D {\bf 96}, no. 6, 064004 (2017)
  [arXiv:1706.07460 [gr-qc]].
\bibitem{6} B. Kleihaus, J. Kunz, S. Mojica and E. Radu, Phys. Rev.D93, no. 4, 044047 (2016) [arXiv:1511.05513 [gr-qc]].
\bibitem{7} D. Ayzenberg and N. Yunes, Phys. Rev. D90, 044066(2014) Erratum: [Phys. Rev. D91, no. 6, 069905 (2015)] [arXiv:1405.2133 [gr-qc]].
\bibitem{8} A. Maselli, P. Pani, L. Gualtieri and V. Ferrari, Phys.Rev. D92, no. 8, 083014 (2015) [arXiv:1507.00680 [gr-qc]].
\bibitem{Nampalliwar:2018iru}
  S.~Nampalliwar, C.~Bambi, K.~Kokkotas and R.~Konoplya,
  Phys.\ Lett.\ B {\bf 781}, 626 (2018)
  [arXiv:1803.10819 [gr-qc]].
\bibitem{13} H. Zhang, M. Zhou, C. Bambi, B. Kleihaus, J. Kunz and E. Radu, Phys. Rev. D95, no. 10, 104043 (2017) [arXiv:1704.04426 [gr-qc]].
\bibitem{14} A. Maselli, L. Gualtieri, P. Pani, L. Stella and V. Ferrari, Astrophys. J.801, no. 2, 115 (2015) [arXiv:1412.3473 [astro-ph.HE]].
\bibitem{15} Z. Younsi, A. Zhidenko, L. Rezzolla, R. Konoplya and Y. Mizuno, Phys. Rev. D94, no. 8, 084025 (2016) [arXiv:1607.05767 [gr-qc]].
\bibitem{16} P. V. P. Cunha,  C. A. R. Herdeiro,  B. Kleihaus, J. Kunz and E. Radu, Phys. Lett. B768, 373 (2017) [arXiv:1701.00079 [gr-qc]].
\bibitem{17} J. L. Blazquez-Salcedo, C. F. B. Macedo, V. Cardoso, V. Ferrari, L. Gualtieri, F. S. Khoo, J. Kunz and P. Pani, Phys. Rev. D94, no. 10, 104024 (2016) [arXiv:1609.01286[gr-qc]].
\bibitem{18} P. Pani and V. Cardoso, Phys. Rev. D79, 084031 (2009)
[arXiv:0902.1569 [gr-qc]]
\bibitem{19} J. L. Blazquez-Salcedo,  F. S. Khoo and J. Kunz, arXiv:1706.03262 [gr-qc].
\bibitem{Konoplya:2018arm}
  R.~A.~Konoplya, Z.~Stuchlík and A.~Zhidenko,
  Phys.\ Rev.\ D {\bf 97}, no. 8, 084044 (2018)
  [arXiv:1801.07195 [gr-qc]].
\bibitem{Konoplya:2010vz}
  R.~A.~Konoplya and A.~Zhidenko,
  Phys.\ Rev.\ D {\bf 82}, 084003 (2010)
  [arXiv:1004.3772 [hep-th]].
\bibitem{Rizzo:2006uz}
  T.~G.~Rizzo,
  Class.\ Quant.\ Grav.\  {\bf 23}, 4263 (2006)
  [hep-ph/0601029].
\bibitem{Cardoso:2008bp}
  V.~Cardoso, A.~S.~Miranda, E.~Berti, H.~Witek and V.~T.~Zanchin,
  Phys.\ Rev.\ D {\bf 79}, 064016 (2009)
  [arXiv:0812.1806 [hep-th]].
\bibitem{Konoplya:2017wot}
  R.~A.~Konoplya and Z.~Stuchlík,
  Phys.\ Lett.\ B {\bf 771}, 597 (2017)
  [arXiv:1705.05928 [gr-qc]].
\bibitem{Ohashi:2004wr}
  A.~Ohashi and M.~a.~Sakagami,
  Class.\ Quant.\ Grav.\  {\bf 21}, 3973 (2004)
  [gr-qc/0407009].
\bibitem{Konoplya:2004wg}
  R.~A.~Konoplya and A.~V.~Zhidenko,
  Phys.\ Lett.\ B {\bf 609}, 377 (2005)
  [gr-qc/0411059].
\bibitem{Konoplya:2005hr}
  R.~A.~Konoplya,
  Phys.\ Rev.\ D {\bf 73}, 024009 (2006)
  [gr-qc/0509026].
\bibitem{Konoplya:2006br}
  R.~A.~Konoplya and A.~Zhidenko,
  Phys.\ Rev.\ D {\bf 73}, 124040 (2006)
  [gr-qc/0605013].
\bibitem{Blazquez-Salcedo:2017txk}
  J.~L.~Blazquez-Salcedo, F.~S.~Khoo and J.~Kunz,
  Phys.\ Rev.\ D {\bf 96}, no. 6, 064008 (2017)
  [arXiv:1706.03262 [gr-qc]].
\bibitem{Gleiser:2005ra}
  R.~J.~Gleiser and G.~Dotti,
  Phys.\ Rev.\ D {\bf 72}, 124002 (2005)
  [gr-qc/0510069].
\bibitem{Takahashi:2012np}
  T.~Takahashi,
  PTEP {\bf 2013}, 013E02 (2013)
  [arXiv:1209.2867 [gr-qc]].
\bibitem{Takahashi:2010gz}
  T.~Takahashi and J.~Soda,
  Prog.\ Theor.\ Phys.\  {\bf 124}, 711 (2010)
  [arXiv:1008.1618 [gr-qc]].
\bibitem{Konoplya:2017zwo}
  R.~A.~Konoplya and A.~Zhidenko,
  JHEP {\bf 1709}, 139 (2017)
  [arXiv:1705.07732 [hep-th]].
\bibitem{Konoplya:2017lhs}
  R.~A.~Konoplya and A.~Zhidenko,
  JCAP {\bf 1705}, no. 05, 050 (2017)
  [arXiv:1705.01656 [hep-th]].
\bibitem{Cuyubamba:2018jdl}
  M.~A.~Cuyubamba, R.~A.~Konoplya and A.~Zhidenko,
  Phys.\ Rev.\ D {\bf 98}, no. 4, 044040 (2018)
  [arXiv:1804.11170 [gr-qc]].
\bibitem{Shinkai:2017xkx}
  H.~A.~Shinkai and T.~Torii,
  Phys.\ Rev.\ D {\bf 96}, no. 4, 044009 (2017)
  [arXiv:1706.02070 [gr-qc]].
\bibitem{Konoplya:2006rv}
  R.~A.~Konoplya and A.~Zhidenko,
  Phys.\ Lett.\ B {\bf 644}, 186 (2007)
  [gr-qc/0605082].
\bibitem{Zinhailo:2018ska}
  A.~F.~Zinhailo,
  Eur.\ Phys.\ J.\ C {\bf 78}, no. 12, 992 (2018).
  [arXiv:1809.03913 [gr-qc]].
\bibitem{reviews}
  R.~A.~Konoplya and A.~Zhidenko,
  Rev.\ Mod.\ Phys.\  {\bf 83}, 793 (2011)
  [arXiv:1102.4014 [gr-qc]];
  E.~Berti, V.~Cardoso and A.~O.~Starinets,
  Class.\ Quant.\ Grav.\  {\bf 26}, 163001 (2009)
  [arXiv:0905.2975 [gr-qc]];
  K.~D.~Kokkotas and B.~G.~Schmidt,
  Living Rev.\ Rel.\  {\bf 2}, 2 (1999)
  [arXiv:gr-qc/9909058].
\bibitem{WKBorder}
B.~F.~Schutz and C.~M.~Will Astrophys.\ J.\ Lett {\bf 291} L33 (1985);
S.~Iyer and C.~M.~Will Phys.\ Rev.\  D {\bf 35} 3621 (1987);
  R.~A.~Konoplya,
  Phys.\ Rev.\  D {\bf 68}, 024018 (2003)
  [arXiv:gr-qc/0303052];
  J.\ Phys.\ Stud.\  {\bf 8}, 93 (2004).
\bibitem{Konoplya:2017tvu}
  R.~A.~Konoplya and A.~Zhidenko,
  Phys.\ Rev.\ D {\bf 97}, no. 8, 084034 (2018)
  [arXiv:1712.06667 [gr-qc]].
\bibitem{Matyjasek:2017psv}
  J.~Matyjasek and M.~Opala,
  Phys.\ Rev.\ D {\bf 96}, no. 2, 024011 (2017)
  [arXiv:1704.00361 [gr-qc]].
\bibitem{Zhidenko-private} A. Zhidenko, et. al. work in progress (2018).
\bibitem{Gundlach:1993tp}
  C.~Gundlach, R.~H.~Price and J.~Pullin,
  Phys.\ Rev.\ D {\bf 49}, 883 (1994)
  [arXiv:gr-qc/9307009].
\bibitem{Kanti:2004nr}
  P.~Kanti,
  Int.\ J.\ Mod.\ Phys.\ A {\bf 19}, 4899 (2004)
  [hep-ph/0402168].
\bibitem{Hawking:1974sw}
  S.~W.~Hawking,
  Commun.\ Math.\ Phys.\  {\bf 43}, 199 (1975)
  Erratum: [Commun.\ Math.\ Phys.\  {\bf 46}, 206 (1976)].
\bibitem{Page:1976df}
  D.~N.~Page,
  Phys.\ Rev.\ D {\bf 13}, 198 (1976).
\end{thebibliography}
\end{document}